# Manipulating the Glass Transition in Nanoscale

D. Y. Sun[a,b] and X. G. Gong[b,c]

[a]School of Physics and Electronic Science, East China Normal University, No.500, Dongchuan Road, Shanghai 200241, People's Republic of China
[b]Key Laboratory for Computational Physical Sciences (MOE), State Key Laboratory of Surface Physics, Department of Physics, Fudan University, Shanghai 200433
[c]Collaborative Innovation Center of Advanced Microstructures, Nanjing 210093

## Abstract

The intrinsic nature of glass states or glass transitions has been a mystery for a long time. Recently, more and more studies tend to show that a glass locates at a specific potential energy landscape (PEL). To explore how the flatness of the PEL related to glass transition, we develop a method to adjust the PEL in a controllable manner. We demonstrate that a relatively flat PEL is not only necessary but also sufficient for the formation of a nanoscale glass. We show that: (1) as long as a nanocluster is located in a region of PEL with local minimum deep enough, it can undergo a first-order solid-liquid phase transition; and (2) if a nanocluster is located in a relatively flat PEL, it can undergo a glass transition. All these transitions are independent of its structure symmetry, order or disorder. Our simulations also uncover the direct transition from one potential energy minimum to another below the glass transition temperature, which is the consequence of flat PELs.

# Introduction

Potential energy landscapes (PELs) provide a unique perspective and indispensable physical picture for exploring the nature of glass and glass transitions.[1-6] The concept of a PEL comes from Goldstein's seminal paper, in which he made a direct connection between glass transitions and PELs.[7] Later, by introducing the inherent structure, Stillinger and Weber further developed the concept of a PEL and established a statistical mechanics framework for a quantitative calculation of thermodynamic properties of glasses.[8] In the past half century, the PEL of some glasses has been investigated.[9-45] Scientists are trying to correlate the properties of glasses with the PEL. For example, Scientists have explicitly proposed a connection between the topology of the PEL (namely the density of configurational states) and the fragility of the associated liquid.[46-48] One breakthrough could be the recognition that the glass transition is rooted in a specific PEL. The PEL of glasses is made up of many metabasins, which are separated by higher barriers.[49-54] In a metabasin, the PEL is relatively flat, even below the glass transition temperature, the system has enough probability jumping from one configuration to another. It was suggested that β relaxation relates to the atomic motion in a metabasin, and α relaxation is the result of inter-metabasin motion.[55,56] With metabasins, the excess configuration entropy, as well as Adam-Gibbs's chain-like motions,[57] may be given a reasonable explanation.[58,59]

With the help of the PEL viewpoint, it seems that we are approaching understanding the nature of glasses. The straightforward way to address these problems is to investigate how thermodynamic behavior changes with variations of the PEL. However, it is not easy to adjust PELs, especially in a controlled manner. There have been a few attempts to study the glass state by adjusting PELs. In these studies, PELs were mostly manipulated by simply changing one parameter of model potentials.[60-66] One obvious disadvantage of the method lies in that the relationship between PELs and the potential parameters is elusive. Especially since the PEL is a complex function in high-dimensional phase space, it seems impossible to establish this kind of relationship, even though previous studies still presented fruitful and instructive results.

In order to establish a relationship between PEL and glass transitions as direct as possible, we propose a controlled manner to adjust 3N-dimensional PELs, where N is the number of atoms. By directly adjusting 3N-dimensional PELs, we have systematically studied the glass transition in a few nanoclusters, both ordered and disordered.

For a long time, there are two parallel themes in the study of glasses, namely bulk glasses and nanoscale glasses. These two themes mutually support and complement, providing different but very important perspectives for exploring the nature of glasses. In addition to its own scientific and technological importance, studies on nanoscale glass has many advantages. First, as demonstrated previously, some nanoclusters can be considered to be the so-called ideal glass, *i.e.*, the glass transition in these nanoclusters can occur at any arbitrarily slow cooling rate[67,68]. This point allows us to study the thermodynamic behavior of glass transitions independent of the cooling rate. Second, for nanoclusters containing dozens of atoms, it is easy to extend the simulation time to microseconds or even longer. It becomes possible to explore the typically slow dynamics of glasses around the glass transition temperature. These two advantages are not available in bulk glass. [69-71]

In this work, we establish the direct relationship between PELs and the glass transition in nanoscale, and found that an enough flat PEL is the key feature for a glass transition, while the symmetry is not direct issue.

## Methodological Development

In order to adjust the 3N-dimensional PEL, we propose the following approach. Assuming $\varphi(r_1, r_2, ... r_N)$ is the total potential energy of a system. In most molecular dynamics (MD) simulations, $\varphi(r_1, r_2, ... r_N)$ is usually expressed as a sum over specific interatomic potentials. Obviously $\varphi(r_1, r_2, ... r_N)$ defines the 3N-dimensional PEL of systems.

We define a new potential energy $\varphi^*$ as,
$$\varphi^* = \begin{cases} \varphi + \varepsilon(\varphi - \varphi_0)^m \ for \ \varphi \leq \varphi_0 \\ \varphi \ \ for \ \varphi > \varphi_0 \end{cases}, \quad (1)$$
where $\varepsilon, \varphi_0$ are adjustable parameters and $m$ is an even number ($m \geq 2$). Thus $\varphi^*$ defines a new 3N-dimensional PEL.

It is easy to calculate the PEL difference between determined by $\varphi^*$ and $\varphi$ in the phase space by directly using Eq. 1. Since
$$\frac{\partial \varphi^*}{\partial r_i} = (1 + \varepsilon m(\varphi - \varphi_0)^{m-1})\frac{\partial \varphi}{\partial r_i} \quad (2)$$
we see that all extreme and saddle points of $\varphi$ pass to $\varphi^*$, keeping their locations in phase space unchanged. In addition, $\varphi^*$ has new extreme points or saddle points determined by
$$1 + \varepsilon m(\varphi_s - \varphi_0)^{m-1} = 0. \quad (3)$$
As we show below, one can carefully select the parameters ($\varphi_0, m, \varepsilon$), to ensure these additional extreme or saddle points are far away from the temperature region where we are interesting.

The PEL described by Eq. 1 is adjusted in the following two aspects. First, suppose $\varphi_a$ and $\varphi_b$ ($\varphi_a < \varphi_b \leq \varphi_0$) being adjacent minimum and maximum points at the PEL of $\varphi$ respectively, and $\varphi_a^*$ and $\varphi_b^*$ being the counterparts at the PEL of $\varphi^*$, we have $\varphi_b^* - \varphi_a^* = \varphi_b - \varphi_a + \varepsilon(\varphi_b - \varphi_0)^m - \varepsilon(\varphi_a - \varphi_0)^m$. Thus if $\varepsilon$ is negative (positive), the barrier in PEL becomes higher (lower). Second, from the second derivatives of the potential energy with respect to atomic positions, namely

$$\frac{\partial^2 \varphi^*}{\partial r_i r_j} = \frac{\partial^2 \varphi}{\partial r_i \partial r_j} + \varepsilon m(\varphi - \varphi_0)^{m-1} \frac{\partial^2 \varphi}{\partial r_i \partial r_j} + (\varepsilon m(m-1)(\varphi - \varphi_0)^{m-2}) \frac{\partial \varphi}{\partial r_i} \frac{\partial \varphi}{\partial r_j}.$$

At any extreme point or saddle point, *i.e.*, $\frac{\partial \varphi}{\partial r_i} = 0 \; or \; \frac{\partial \varphi}{\partial r_j} = 0$, we have

$$\frac{\partial^2 \varphi^*}{\partial r_i r_j} = \frac{\partial^2 \varphi}{\partial r_i r_j}(1 + \varepsilon m(\varphi - \varphi_0)^{m-1}). \quad (4)$$

one can see that, compared to that of $\varphi$, the PEL of $\varphi^*$ at extreme points can be adjusted to be more or less flat, based the choice of $\varepsilon$, $\varphi_0$ and *m*. For example, if $\frac{\partial^2 \varphi}{\partial r_i r_j} > 0$, by taking $(1 + \varepsilon m(\varphi - \varphi_0)^{m-1}) > 0$ the PEL of $\varphi^*$ will become steeper. From the above two points, we can see that, we are able to not only directly adjust the height of the barrier (the first point), but also adjust the second derivative of the PEL (the steepness of PELs) near the extreme point (the second point).

It needs to point out that, by adjusting the flatness of PELs, other than the additional extreme or saddle points determined by Eq. 3, we do not change either the number or position of extreme points in phase space, which is a remarkable feature of our method. This statement is easy to see from equation 2, namely, except the additional extreme points determined by Eq. 3, as long as the first derivative of $\varphi$ is zero, does the first derivative of $\varphi^*$. We note that, similar ideas have been used to study the dynamics of polymers.[72-74] as a mechanism of accelerating molecular dynamic simulations.

Using MD simulations, we have studied the melting behavior of two aluminum nanoclusters ($Al_{43}$ and $Al_{55}$). For Al, the interatomic potential is adopted the glue potential[75](**Note: the total potential energy ($\varphi$) discussed above is not the interatomic potential but a kind of sum of it**). It is known that, at $\varepsilon=0$ $Al_{43}$ has a disordered ground state structure.[67] The ground state of $Al_{55}$ has an ordered structure with high symmetry ($I_h$ symmetry).[76] The melting of $Al_{55}$ is a typical solid-liquid phase transition, while $Al_{43}$ melts and solidifies with a typical glass-like transition, which has been suggested to be an ideal glass.[68]

In current studies, for $Al_{43}$, $\varphi_0$ and *m* are -2.58× 43 eV and *6*, respectively, and $\varepsilon$ is chosen in the range of $[-4 \times 10^{-5}, 0]$. Since here $\varepsilon$ is negative, the PEL becomes steeper as $\varepsilon$ is decreased. For $Al_{55}$, $\varphi_0$=-2.58× 55 eV, *m=4*, and $\varepsilon$ is in the range [0.0, $2.8 \times 10^{-4}$]. Since here $\varepsilon$ is positive, the PEL becomes flatter as $\varepsilon$ is increased. Our simulations confirm that for all chosen $\varepsilon$, the disordered structure and $I_h$ structure are still the stable configurations for $Al_{43}$ and $Al_{55}$, respectively.

For both nanoclusters, $\varphi_0$=-2.58 eV times the number of atoms, corresponds to the total potential energy above the melting temperature of $\varepsilon=0$. It is easy to calculate that the additional extreme or saddle points determined by Eq. 3 are far away from the potential range of interest. The value of $\varphi_s$ is easily gotten from Eq. 3. For $Al_{43}$, $\varphi_s > \varphi_0$, and the corresponding $\varphi_s$ for all $\varepsilon$ is higher than 900 K, which is irrelevant to any phase or glass transition in the current studies. For Al$_{55}$, $\varphi_s < \varphi_0$, for all studied $\varepsilon$, and the corresponding temperature is less than ~200K, which is much lower than the melting point or glass transition temperatures. In fact, the extra extreme or saddle points at $\varphi_s$ have essentially no effect on the thermodynamic properties.

## Result and Discussion

The basic difference between a glass transition and a solid-liquid phase transition lies in how the energy and volume change. In a glass transition process, both energy and volume change continuously over the whole range of temperatures. While in a solid-liquid phase transition, both energy and volume have a jump at the melting temperature.[77] Figs. 1 and 2 depict the change of energies, volumes and specific heats with temperature for different PELs adjusted by different parameters. The melting behavior is closely related to the flatness of the PEL regardless of the structural symmetry. For the disordered Al$_{43}$, with the increase of ε, for which the PEL becomes steeper and steeper, the melting behavior changes accordingly from a typical glass transition to a first order solid-liquid phase transition. For the ordered Al$_{55}$, when the PEL gets progressively flatter by increasing ε, the melting behavior changes from a typical first order solid-liquid phase transition to a glass transition. These results indicate that the glass transition is not intrinsically dependent on the structural symmetry, but on the flatness of a PEL. When the PEL becomes steeper, a first order solid-liquid phase transition occurs; when the PEL becomes flatter, a glass transition occurs.

At ε=0, the melting of Al$_{43}$ is a typical glass transition, indicated by the continuous change in energy (black circles in upper panel of Fig. 1) and volume (black circles in upper panel of Fig. 2), as we have demonstrated several times in our previous work.[78,79] The glass transition temperature, which is around 520K, can be estimated by a linear extrapolation of the low temperature and high temperature dependence of energies. As $\varepsilon$ decreases from zero to negative, which means the PEL gets steeper and steeper, the melting of Al$_{43}$ shows a typical first order solid-liquid phase transition of finite systems, as shown in Fig. 1 (upper panel). For ε=-2 × $10^{-5}$, the melting of Al$_{43}$ has begun to deviate from the glass transition behavior. For ε=-4 × $10^{-5}$, the energy changes rapidly, and a significant latent heat appears. The step around melting is not very sharp, which is a remarkable feature of the finite size effect of melting.[80] The clear fact beyond question is that a typical first-order phase transition does occur in this disordered glass-like Al$_{43}$ after the PEL is adjusted, *i.e.*, if the PEL of glass becomes deeper, it shows a typical solid-liquid transition.

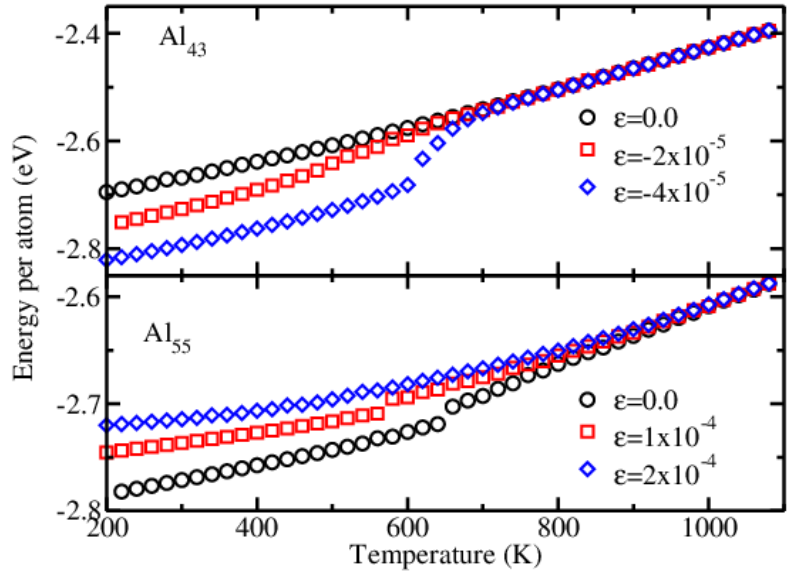

**Figure 1: (Color online)** The energy as a function of temperature for ordered $Al_{55}$ (lower panel) and disordered $Al_{43}$ (upper panel). For $Al_{43}$, with the increase of ε, the continuous change in energy is gradually replaced by a step, while for $Al_{55}$, with the increase of |ε|, a step in energy is gradually replaced by a continuous change.

The lower panel of Fig. 1 shows the energy as a function of temperature for the ordered $Al_{55}$ with a range of ε. The energy has a clear step for ε=0.0 and $1 \times 10^{-4}$ at the melting temperature, a typical feature of a first-order phase transition. However, for ε=$2 \times 10^{-4}$, the energy step disappears, resulting in a typical glass transition process. This indicates that, if an ordered nanocluster is trapped in a relatively flat PEL, a glass transition can occur.

The change of volume with temperature further supports our conclusions drawn from energy changes. Fig. 2 shows the volume as a function of temperature for $Al_{43}$ (upper panel) and $Al_{55}$ (lower panel). For the disordered $Al_{43}$, when ε=0, the volume changes continuously with typical glass transition characteristics. When ε=-$5 \times 10^{-5}$, the volume jumps obviously at the melting point, showing characteristics typical of a solid-liquid phase transition in finite systems. For ordered $Al_{55}$, the volume clearly jumps for ε=0 and $1 \times 10^{-4}$ around the melting temperature. This is a typical feature of a first-order phase transition. When ε=$2 \times 10^{-4}$, the step has disappeared, reflecting a typical glass transition process.

The change in melting behavior for different PELs can also be found from specific heat, as shown in Fig. 3. For $Al_{43}$ (lower panel of Fig. 3), for ε=0 the specific heat does not exhibit any peak or discontinuity, indicating a typical glass transition. With ε=-$2 \times 10^{-5}$, the peak in specific heat becomes sharp, which is a distinct feature of melting in finite-size systems. The appearance of a peak instead of a discontinuity is due to the existence of the solid-liquid coexistence for finite-size systems. For $Al_{55}$ (upper panel of Fig. 3), we have identified a change from the first-order solid-liquid phase transition to the glass transition as the PEL getting flatter and flatter. One can see that, for ε=0 and $1 \times 10^{-4}$, the specific heat has a step at the melting point, which is a hallmark of solid-liquid phase transitions. While it becomes a continuous change and does not show any peak in the entire temperature range for ε=$2 \times 10^{-4}$, implying a glass transition.

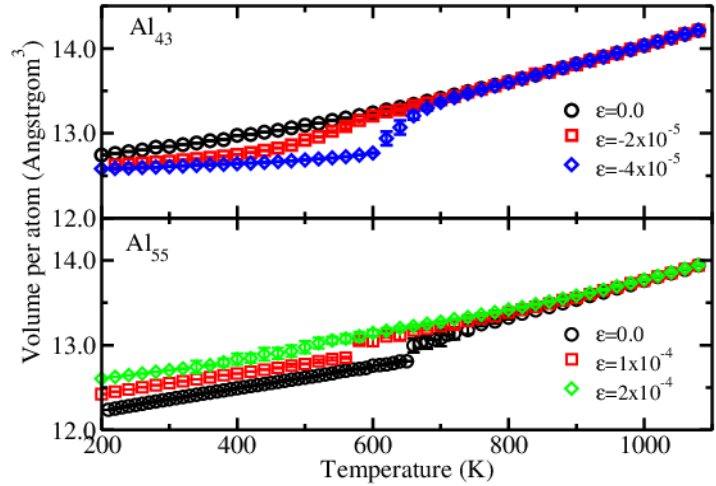

**Figure 2: (Color online) The volume as a function of temperature for ordered Al$_{55}$ (lower panel) and disordered Al$_{43}$ (upper panel). The change in volume is similar to that shown in Fig. 1.**

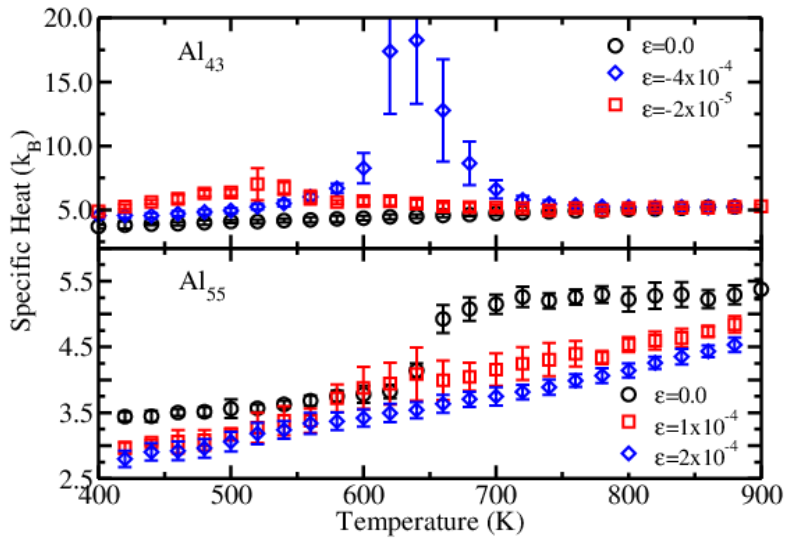

**Figure 3: (Color online) The specific heat as a function of temperature for disordered nanocluster Al$_{43}$ (lower panel) and ordered nanocluster Al$_{55}$ (upper panel). For Al$_{43}$, with the decrease of ε, $i.e.$, the PEL becomes deeper, the peak in specific heat becomes sharper and sharper, indicating evolution from a glass transition to a first order phase transition. For Al$_{55}$, with the increase of |ε|, $i.e.$, the PEL becomes flatter, the jump in specific heat disappears gradually, indicating evolution from a first order phase transition to a glass transition.**

Principal radii of gyration, in addition to indicating a change of nanocluster shape, also identify the type of phase transition. We find that the difference between the maximum and minimum radius of gyration ($\Delta R=R_{max}-R_{min}$) gives clear information about structural changes and melting behavior. Fig. 4 shows $\Delta R$ as a function of temperature. For a first order solid-liquid phase transition ($Al_{43}$ with $\varepsilon=-3.4 \times 10^{-5}$, $Al_{55}$ with $\varepsilon=1.0 \times 10^{-4}$), the nanoclusters keep their original structure before completely melting, indicated by an approximately constant value of $\Delta R$. A jump in $\Delta R$ around the melting temperature can be clearly seen for these nanoclusters. For a typical glass transition ($Al_{43}$ at $\varepsilon=-0.4 \times 10^{-5}$, $Al_{55}$ at $\varepsilon=2.0 \times 10^{-4}$), far below $T_g$, $\Delta R$ has already begun to change. With the increase of temperature, in contrast to the first solid-liquid phase transition, $\Delta R$ gradually changes, spanning temperatures over a range of about 200K. We can define the temperature at which nanoclusters begin to change shape as the starting temperature ($T_s$), which represents a characteristic temperature, above which configurational entropies emerge in glassy states.

According to Fig. 4, three characteristic temperatures, $T_s$, $T_g$ and $T_m$, can be determined. $T_m$ is the obvious one without any ambiguity. It is defined as the temperature, at which $\Delta R$ jumps, which is in accordance with abrupt changes in the volume and energy as shown in Figs. 1 and 2. $T_s$ can be easily defined as the temperature at which $\Delta R$ obviously begins to change. In contrast, it is not easy to determine $T_g$, which is a well-known difficulty. However, Fig. 4 shows us that the glass transition is much more clearly reflected on the $\Delta R$ -T curve. In the current work, $T_g$ can be estimated by linear extrapolation of the low temperature and high temperature dependence of $\Delta R$, as illustrated by dash lines in Fig. 4. Two issues need to be emphasized: 1) $T_g$ is consistent with that determined by other thermodynamic quantities, energies, volumes, *etc*; and 2) It is inevitable that there can be large errors, as is also the case in other methods.

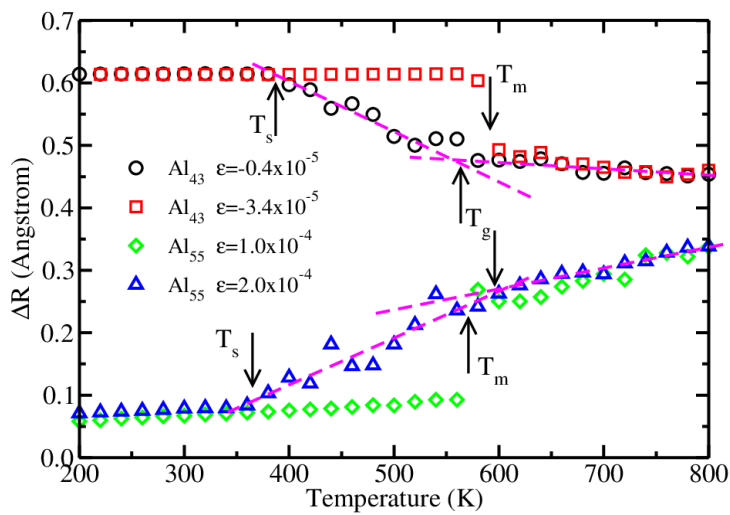

**Figure 4. (Color online) The difference between the maximum and minimum radius of gyration.**

There seems to be a consensus that cooperative diffusion within a few atoms exists in a glass. This perspective comes from Adam-Gibbs entropy theory and Goldstein's PEL picture.[7,57] Recently, it has become generally believed that this cooperative diffusion may correspond to a transition in a metabasin. We show that the atomic diffusion in a glass below $T_g$ can indeed be considered as a transition from one potential energy minimum to another, for which the transition is accomplished by cooperative diffusion.[81-85].

We have calculated the diffusion coefficient, which is shown in Fig. 5 for $Al_{43}$ (upper panel) and $Al_{55}$ (lower panel). If nanoclusters melt through a solid-liquid phase transition, the atomic diffusivity becomes negligible below the melting point, indicating a normal solid state. For both $Al_{43}$ and $Al_{55}$, if a glass transition occurs, the atoms have obvious diffusivity below $T_g$. This is a typical feature for glassy nanoclusters suggested recently.[78,79]

The diffusion activation energy can be obtained by fitting the diffusion coefficient via temperatures. For glass transitions, diffusion activation energies are clearly different below and above $T_g$. In particular, the glass state (low temperature) has lower activation energy than that in the liquid state (high temperature), in agreement with previous studies.[86,87] As we have pointed out previously, this is a typical characteristic of diffusion in glasses.[78,79] It is well known that the diffusion of atoms in liquids is mainly determined by thermal collisions. The change in activation energy around $T_g$ implies a change of diffusion mechanism. As discussed in detail in our previous works, this diffusion is a kind of collective diffusion, and the diffusion barrier is even lower than that in the liquid.[79]

Obviously, this cooperative diffusion is the result of a relatively flat PEL, because it only occurs with a sufficiently flat PEL, *i.e.*, $\varepsilon \geqslant -1 \times 10^{-5}$ for $Al_{43}$ and $\varepsilon \geqslant 2 \times 10^{-4}$ for $Al_{55}$. One can conclude that the relatively flat PEL is the essence of the glassy state, and results in cooperative diffusion at low temperature.

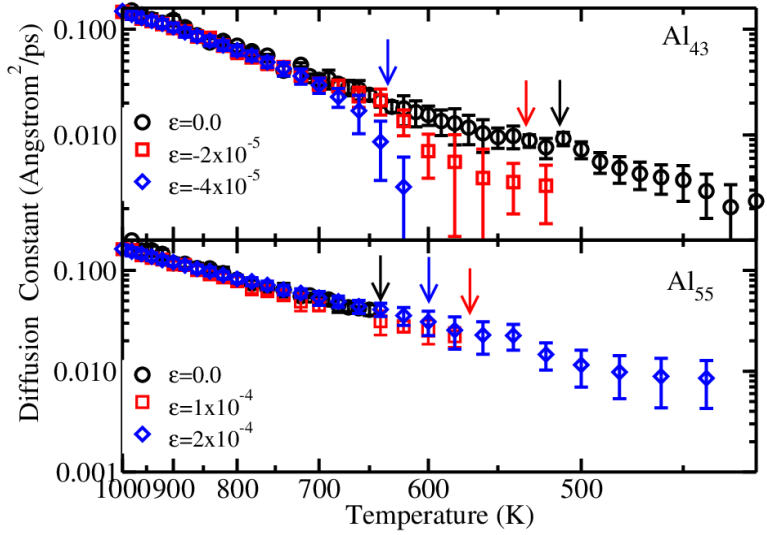

**Figure 5.** (Color online) Temperature dependence of the diffusion constant for $Al_{43}$ (upper panel) and $Al_{55}$ (lower panel), where the vertical axis is logarithmic and the horizontal axis is reciprocal scale. For $Al_{43}$ at $\varepsilon=0.0$ and $Al_{55}$ at $\varepsilon=2\times 10^{-4}$, there is the evident diffusivity below the glass transition temperature. The arrows, with the same color as data points, roughly indicate the corresponding $T_g$ or $T_m$.

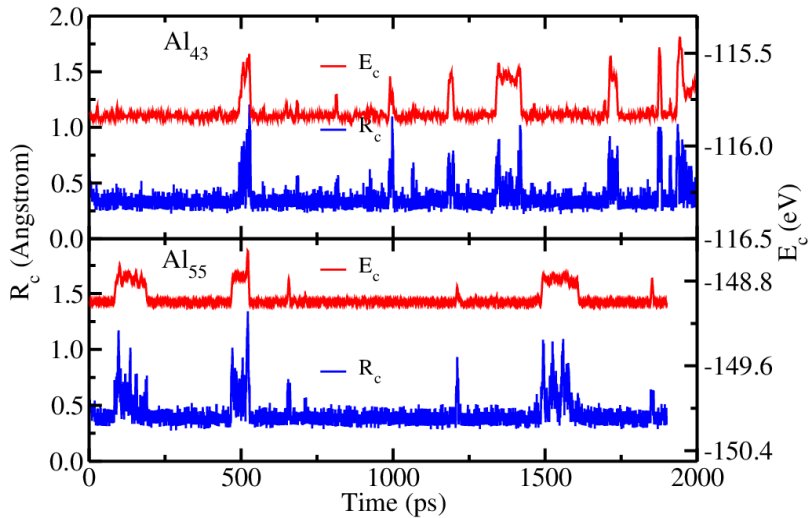

**Figure 6.** (Color online) The short-time average of potential energy ($E_c$) and displacement ($R_c$) for $Al_{43}$ at $\varepsilon=0.0$ (upper panel) and $Al_{55}$ at $\varepsilon=2\times 10^{-4}$ (lower panel), where the temperature is 400K.

This cooperative diffusion corresponds to a jump from one local minimum to a neighboring local minimum, still in the relatively flat PEL. The direct way to show this is to calculate the evolution of total potential energies and atomic displacements. In order to separate the thermal fluctuation from the energy and atomic displacements, the short-time average of the energy ($E_c(t)$) and atomic displacement ($R_c(t)$) are calculated by averaging over a certain time interval. More concretely, the short-time average of displacement is

$$R_c(t) = (\frac{1}{N}\sum_{i=1}^{N}(r_i(t) - r_i(t+\Delta t))^2)^{\frac{1}{2}},$$

where N is the number of atoms, $r_i(t)$ denotes the position of the *i*th atom at time t, and $\Delta t$ is a short duration, 5 ps in current studies. Our previous study has shown that $R_c(t)$ can be used to identify the cooperative motion.[78] The short-time average of the total potential energy is

$$E_c(t) = \frac{1}{\Delta t}\int_0^{\Delta t} E(t+\tau)d\tau,$$

where E(t) is the instantaneous total potential energy, and Δt is again 5 ps. If $E_c(t)$ takes a certain value at one time period and is a different value during another time period, the system can be considered to be at different local minimum of the PEL. If $E_c(t)$ is synchronized with a notable change in $R_c(t)$, we can conclude that the cooperative diffusion corresponds to the jump from one PEL minimum to another.

Fig. 6 shows the evolution of $R_c$ and $E_c$ over time. The upper panel of Fig. 6 corresponds to $Al_{43}$ at ε = 0.0, the lower panel of Fig.6 corresponds to $Al_{55}$ at ε = $2 \times 10^{-4}$, both at T=400K. Both $E_c$ and $R_c$ maintain relatively small values over many time periods, which corresponds to one or more equilibrium states with similar energy. However, in some short periods of time, $R_c$ has a rapid increase. By carefully observing the atomic trajectory, we find that the rapid increase of $R_c$ is always associated with a larger displacement of a few atoms at the same time. The size of displacement is comparable to the average distance between atoms, indicating that diffusion occurs. This process can be considered as a cooperative diffusion. At the same time, a rapid increase of $E_c$ occurs simultaneously, which should relate to the activation process of this cooperative diffusion. After a rapid increase, $R_c$ quickly returns to the smaller value, then $E_c$ equilibrates to a new state. Perhaps it corresponds to a transition from one inherent structure to another. Since the two equilibrium states are very close in energy, which is precisely what people speculate or assume about a metabasin. Although only a few cooperative diffusion events are shown, such behaviors are common in glassy states.

Combining the current result with our previous work,[78] we can present a general phase behavior of a glass. Thermodynamically, a solid, glass and liquid mainly differ in two aspects, internal energies and configurational entropies. The difference in energies is obvious and easily understood. However, the difference in configurational entropies is ambiguous. A solid is in a deep potential well, in which the vibrational entropy is important, while the configurational entropy is neglectable. In liquids, the atoms diffuse quickly, the configurational entropy reaches a maximum compared to solids and glasses. In glasses, the atom vibrates at a local potential minimum for most of the time. As we have demonstrated, the remarkable feature of glasses is the existence of collective diffusion, thus the configurational entropy should play a role on the thermodynamical behavior of glasses. the measured vibrational entropies of the glass and liquid show a tiny excess over the crystal, representing less than 5% of the total excess entropy measured with step calorimetry. A recent experiment has also shown that the excess entropy of metallic glasses is almost entirely configurational in origin.[88]

By adjusting the PEL, we are able to make some comprehensive comparisons. Fig. 7 shows the change of $T_s$, $T_g$ and $T_m$ as a function of $\varepsilon$. It can be seen that, $T_g$, for either $Al_{43}$ or $Al_{55}$, shows a significant jump when the melting behavior of the nanocluster changes from a first-order solid-liquid phase to a glass transition. This jump should be caused by the emergence of configuration entropies.

Finally, the last issue need to be emphasized regarding to current results. Different from bulk glasses, the glass presented in this paper should be identified as the ideal glass. Because in our simulations, the system is in complete equilibrium at each temperature, guaranteed by the much long equilibrium time up to microseconds. It means that the cooling rate tends to be infinite, which is the theoretical condition for obtaining ideal glass. Usually, in the vitrification of bulk liquid, it is impossible to achieve truly infinite slow cooling rate, because as the cooling rate is lower than a critical value, the crystallization will occur before the glass transition. In this paper, the existence of these ideal glasses should be related to two folds. First, due to the lack of translation symmetry, the disordered structure can be the ground state of nanoclusters. In fact, as early as 20 years ago, the current authors found the glass transition behavior in nanoclusters.[67] The second reason should closely relates to the adjustment on PEL. When PEL is adjusted flat enough, the system is able to escape from the potential well before complete melting, thus the configuration entropy will emerge and leading to the glass transition.

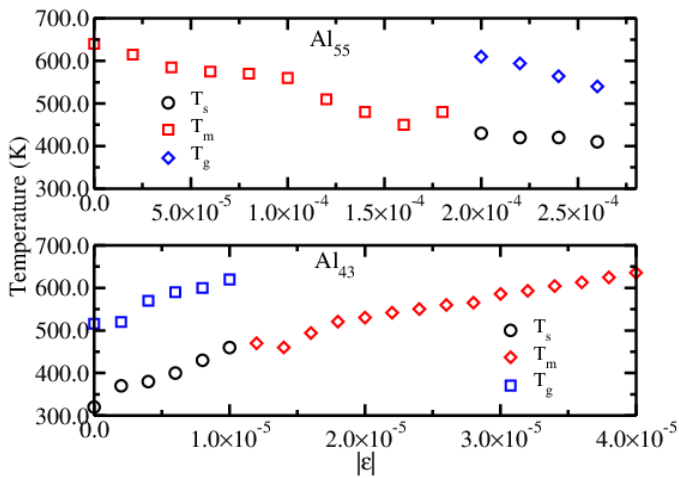

**Figure 7. (Color online) $T_s$, $T_g$ and $T_m$ as a function of $\varepsilon$. For $Al_{43}$, $\varepsilon < -1 \times 10^{-5}$, the disordered nanoclusters begin to exhibit a first-order solid-liquid phase transition. For $Al_{55}$, $\varepsilon > 2 \times 10^{-4}$, the ordered nanoclusters begin to undergo a glass transition.**

## Summary

In this paper, we have systematically studied the glass transition of nanoclusters by molecular dynamics. We have proposed a new method to adjust the flatness of a potential energy landscape. This method can adjust the height of potential barriers without changing the number or position in phase space of the potential energy extremal points. By adjusting the flatness of the potential energy surface, we have found that the nanoclusters can undergo either a first-order solid-liquid phase transition or a glass transition, which is independent of the structural symmetry. This makes it possible for us to demonstrate for the first time that a relatively flat potential energy landscape is an intrinsic nature for glass transitions. Because these nanoclusters undergo an ideal glass transition, *i.e.* under an arbitrarily slow cooling rate, it allows study of the dynamic processes at any long time. We find that, even under the glass transition temperature, the system can frequently transfer from one potential surface minimum to another, and this process is achieved by a collective diffusion, which is completely different from the thermal collision process in liquids. Our study provides, for the first time, a microscopic version of atomic motion below the glass transition temperature, as well as a physical picture of the potential energy landscape.

## Acknowledgements

Project supported by the National Natural Science Foundation of China (Grant No. 11874148). The computations were supported by ECNU Public Platform for Innovation.